\documentclass[twocolumn,           % Format : preprint, twocolumn
               showpacs,            % Pacs : showpacs, noshowpacs
               preprintnumbers,     % Preprint: preprintnumbers,
               aps,                 % Society: ...
               prd,                 % Journal Style : pra, prb, prc, prd, pre,
               superscriptaddress,      % Affiliation (Title) : groupedaddress,
               nofootinbib,         % Footnote: footinbib, nofootinbib
               tightenlines,        % Remove additional spaces in a line
               floats,floatfix      % Floating pictures and tables
               ]{revtex4}

\usepackage{graphicx}
\usepackage{latexsym}
\usepackage{amsmath,amssymb}        % amssymb includes amsfonts
\usepackage[draft=false]{hyperref}

\begin{document}

% -----> TITLE PAGE 

\title{Scalar Field Dark Matter: head-on interaction between two 
structures}

\author{Argelia Bernal}
\affiliation{Departamento de F\'{\i}sica, Centro de
        Investigaci\'on y de Estudios Avanzados del IPN, AP 14-740,07000
        M\'exico D.F., M\'exico.}

\author{F. Siddhartha Guzm\'an}
\affiliation{Instituto de F\'{\i}sica y Matem\'{a}ticas,
        Universidad Michoacana de San Nicol\'as de Hidalgo. Edificio C-3, 
	Cd. Universitaria,
        C. P. 58040 Morelia, Michoac\'{a}n, M\'exico.}

% --->   DATE

\date{\today}

% -----> ABSTRACT

\begin{abstract}
In this manuscript we track the evolution of a system consisting 
of two self-gravitating virialized objects made of a scalar field in the 
newtonian limit. The Schr\"odinger-Poisson system contains a potential 
with self-interaction of the Gross-Pitaevskii type for Bose 
Condensates. Our results indicate that solitonic behavior is allowed in the 
scalar field dark matter model when the total energy of the system is positive, 
that is, the two blobs pass through each other as should happen for solitons; on 
the other hand, there is a true collision of the two blobs when the total energy 
is negative. 
\end{abstract}

% ----->   PACS


% ----->   MAKETITLE   <-----

\maketitle

% -------------------------------------
% 	        ARTICLE
% -------------------------------------

% ----->     INTRODUCTION

\section{Introduction}
\label{sec:introduction}

The most widely studied dark matter hypothesis consists in 
assuming that it is made of point-like cold particles that 
are responsible for the formation of structure in the universe; among the 
most studied candidates nowadays are the supersymmetric particles that 
would behave as a cold fluid made of particles. However, two 
problems associated to the point-like nature of dark matter are that the 
resulting gravitational collapse shows a central density profile that is 
not flat and on the other hand it predicts a non-observed amount of 
small structures. An alternative to ameliorate these two problems 
consists in assuming that the dark matter is made of an ultralight 
spinless particle, the so called Scalar Field Dark Matter Model (SFDM). In 
the cosmological frame, the analysis of such hypothesis indicated that the 
power of structures could be controlled through a parameter in the model, 
that is, the mass of the scalar field representing the spin-less particle 
\cite{varun,MatosUrena2000a,MatosUrena2000b,MatosUrena2004}. Once the mass $m$ of 
the 
boson is 
fixed the power spectrum suffers a cutt-off according to the mass of 
the 
smallest structure desired. An interesting assumption in such analysis is 
that the scalar field potential was a cosh-like potential, that behaved as 
an exponential at early times and as a free field case (quadratic 
potential) at late times, whose behavior was that of the usual cold dark 
matter model. Moreover, it was found that the SFDM enjoys the same 
advantages at cosmic scale as the standard lambda cold dark matter model.

Because the SFDM requires the existence of a fundamental scalar field for 
its reliability, it is natural to consider that this scenario fits very 
well within unification theory scenarios and braneworld models 
\cite{BWconnection}. This by itself is a good enough reason to consider 
the SFDM as an alternative powerful model, because it contains 
intrinsically the spinless boson as dark matter particle. However, once at 
cosmic scales the model matches with observations, it is necessary to 
study the predictions of the model at structure scales. In this sense 
there have been several results indicating that the model is good also 
at galactic scales and here we briefly summarize such results. 

At the early stages of the galactic dark matter model, fully general 
relativistic stationary solutions were proposed to explain the phenomena 
like the flatness of rotation curves, assuming the scalar field was real 
\cite{GuzmanMatos1999,Mbelek}, 
non-topological scalar field dark halos\cite{MielkeSchunk2002}
complex scalar fields\cite{Schunck1998} and global 
monopoles \cite{Nucamendi}. Later on, the assumptions relaxed to the 
newtonian limit of such general relativistic models. By the time, 
quintessential dark matter halos \cite{annalen,Arbey1,Arbey2} and the fluid 
dark matter made of scalar fields was proposed as an alternative galactic 
dark matter model\cite{Arbey3} and the collapse of fuzzy 
dark matter made of a scalar field was analyzed in one 
dimension\cite{Hu2000}. On the other hand, the assumption of time independence was 
also relaxed and scalar field dark matter halos were proposed to be 
gigantic Oscillatons, that is, time dependent fully relativistic scalar 
field solutions to the Einstein-Klein-Gordon system of equations
\cite{galacticolapse,phi2-oscillatons}. 

Currently what appears to be the interesting case is that of the time 
dependent newtonian limit of the model, that is, the 
Schr\"odinger-Poisson (SP) system of equations would describe the model 
at local scales. In this direction relevant results have been found, for 
instance: it was shown that when the evolution of a structure of galactic 
mass is followed after the turnaround point, it quickly virializes and 
tends to a stationary equilibrium solution of the SP system of equations, 
whereas one of the size of a supercluster would still be relaxing at the 
present time\cite{GuzmanUrena2003}; the condition is that the mass of the 
boson ($m \sim 10^{-23}$eV) is the one that better cuts-off the power 
spectrum at galactic scales as shown in 
\cite{varun,MatosUrena2000a,MatosUrena2000b}. Thus, at the moment the pieces 
of the model seem to match both, at cosmic and at local scales. In fact, 
recently in \cite{GuzmanUrena2006} it was shown that the scalar field 
gravitational collapse tolerates the introduction of a self-interaction 
term in the potential, which makes the model to seem quite like a 
self-gravitating Bose-Condensate. In \cite{BernalGuzman2006} we showed 
that spherically symmetric equilibrium solutions of the SP system are 
stable against non-spherical perturbations, and moreover, such 
configurations played the role of late-time attractors for initially quite 
general axisymmetric initial density profiles.

What we present here is a step forward in the direction of studying the
evolution of scalar field structures. We perform numerical studies of 
scalar binary configurations, as a first step towards the making of a 
numerical code with no symmetries and for N-scalar objects. These studies 
would tell us about possible restrictions on self-interaction terms for 
the scalar field, and the way single configurations interact with each
other. We restrict ourselves to the case of head--on interaction, which 
can be handled with a 2D code with axi-symmetry. We choose to write down 
the SP in cylindrical coordinates:

\begin{eqnarray}
i \frac{\partial \psi}{\partial t} &=& -\frac{1}{2} \left(
                                \frac{\partial^{2} \psi}{\partial x^2}
                              + \frac{1}{x}\frac{\partial \psi}{\partial x}
                              + \frac{\partial^{2} \psi}{\partial z^2}
                                \right)
                    + U \psi + \Lambda|\psi|^2\psi\label{eq:schroedinger}\\
\frac{\partial^{2} U}{\partial x^2} &+&
\frac{1}{x}\frac{\partial U}{\partial x} +
\frac{\partial^{2} U}{\partial z^2} =
\psi^{\ast}\psi.\label{eq:poisson}
\end{eqnarray}

\noindent where $\psi=\psi(x,z,t)$ and $U=U(x,z,t)$ are the wave function 
and the gravitational potential respectively; $x,z$ are the radial and 
axial cylindrical coordinates respectively. The third order term in 
Eq.~(\ref{eq:schroedinger}) is related to a self-interacting term, in
which $\Lambda$ corresponds to the s-wave scattering length in the
Gross-Pitaevskii approximation for Bose Condensates 
\cite{PitaevskiiGross}. This term was shown to play the role of 
determining the compactness of the structure \cite{GuzmanUrena2006}. 
Equations (\ref{eq:schroedinger}-\ref{eq:poisson}) use the 
units and scaling $\hbar = c = 1$ with $x\rightarrow mx$, $z\rightarrow mz$, $t 
\rightarrow mt$ and the wave function $\psi \rightarrow \sqrt{4\pi G}\psi$, where 
$m$ is the mass of the ultralight boson. As a consequence of this change of units 
is that the mass of a system will be in units of $[M]=M^{2}_{pl}/m$ as found also 
for fully relativistic boson star solutions 
\cite{Guzman2006,GuzmanUrena2006}; this implies that the 
value $m$ sets the physical lenght and time scale of the configuraions evolved. 
This mass, together with the scaling relations of the Schr\"odinger-Poisson 
system \cite{GuzmanUrena2003,BernalGuzman2006,GuzmanUrena2004} are the basics for 
transforming back to physical units the system of interest (see an exmple below).

The paper is organized as follows. In the next section we briefly describe 
the code used. In section \ref{sec:initialdata} we construct 
initial data containing two spherically symmetric equilibrium 
configurations along the $z$ axis. In section \ref{sec:headon} we show the 
results for the head-on interaction of the two structures. Finally in 
section \ref{sec:conclusions} we draw some conclusions.

% ----->     NUMERICAL METHODS

\section{Numerical methods}
\label{sec:numerical-methods}

{\it The evolution.} The most common numerical technique for
time-integrating Eq.~(\ref{eq:schroedinger}) is implicit with
alternating direction splitting of the evolution
operator \cite{Choi,Harrison}. The reason for this is that the
evolution operator is unitary. Nevertheless, we used such method
in \cite{GuzmanUrena2004}, where no need for splitting the operator on
the right hand side of Eq.~(\ref{eq:schroedinger}) was needed; 
in \cite{GuzmanUrena2006,BernalGuzman2006} it was shown that explicit methods 
preserve also the number of particles and no significant difference in 
the results is found after using one method or the other. For the present 
case an explicit approximation of the full implicit method (in practice, a 
modified iterative Crank-Nicholson method \cite{phi2-oscillatons}), with 
second order finite differencing to calculate the spatial derivatives is 
used. The reason to avoid using the implicit method is the difficulty to 
reduce the evolution operation to a tridiagonal system of equations when 
considering a non-zero $\Lambda$, which makes the Schr\"odinger equation a 
non-linear one, a situation not descussed in \cite{Choi,Harrison}.\\

{\it Poisson equation.} Equation (\ref{eq:poisson}) is an elliptic
equation for $U$ which we solve using the 2D five-point stencil for the
derivatives and a successive over-relaxation (SOR) iterative algorithm
with optimal acceleration parameter (see e.g.\cite{Smith1965} for details
about SOR). In order to impose boundary conditions we made sure the
boundaries were far enough for the mass 
$M=\int |\psi|^2 d^3x$ to be the same along the three faces of the 
domain and used the
monopolar term of the gravitational field; that is, we used the value 
$U = -M/r$ along the boundaries with $r=\sqrt{x^2+z^2}$. At the axis we
demanded the gravitational potential to be symmetric with respect to the
axis.

We use a sponge in the outermost region of the domain. The sponge is a
concept used with success in the past when dealing with the Schr\"odinger
equation (for detailed analyses see \cite{GuzmanUrena2004,Israeli1981}). 
This technique consists in adding up to the potential
in the Schr\"odinger equation an imaginary part. The result is that
in the region where this takes place there is a sink of particles, and
therefore the density of probability approaching this region will be
damped out, with which we get the effects of a physically open
boundary.

Basic testbeds of this code evolving single equilibrium configurations can 
be found in \cite{BernalGuzman2006}, where the results are also compared 
with previous studies with spherical symmetry and linear perturbation 
theory.

% ----->     INITIAL DATA

\section{Initial data}
\label{sec:initialdata}

Details about the construction of initial data for 
spherically symmetric equilibrium configurations can be found in 
\cite{GuzmanUrena2004,GuzmanUrena2006,BernalGuzman2006}. 
Here we briefly mention the procedure used in our binary case. What we do 
is to superpose two spherically symmetric ground state 
equilibrium configurations upon 
the same 2D axially symmetric grid, whose construction is described as 
follows. In spherical symmetry equations 
(\ref{eq:schroedinger},\ref{eq:poisson}) read

\begin{eqnarray}
i\partial_t \psi &=& -\frac{1}{2r} \partial^2_r (r\psi) + U \psi
+ \Lambda |\psi|^2\psi\label{eq:sph_schroedinger}\\
\partial^2_r (rU) &=& r \psi \psi^\ast . \label{eq:sph_poisson}
\end{eqnarray}

\noindent where $r=\sqrt{x^2+z^2}$. If a time dependence of 
the type $\psi = \phi e^{i\omega t}$, regularity at the origin, 
%$\phi(0)=\partial_x \phi(0)=0$ 
and an isolation condition $\phi(r\rightarrow \infty) = 0$ are assumed, the 
system becomes an eigenvalue problem for $\phi$ with eigenvalue $\omega$:

\begin{eqnarray}
\partial^{2}_{r}(r\phi)&=& 2 r (U-\omega) + 2\Lambda
|\phi|^2\phi \, , \label{eq:sph_schroedinger_eq} \\
\partial^{2}_{r}(rU)&=& r\phi^2 \, . \label{eq:sph_poisson_eq}
\end{eqnarray}
                                                                                
\noindent 
In order to solve these equations we discretize them and use a shooting 
method that bisects the value of $\omega$ so that the boundary 
conditions hold with certain desired accuracy. The solutions constructed 
in this way are called equilibrium configurations. For each value of 
$\Lambda$ it is possible to construct the whole branch of equilibrium 
configurations as shown in \cite{GuzmanUrena2006}. However there is an 
extra ingredient in these solutions: the number of nodes of the wave 
function. When the wave function is nodeless we say it belongs to the 
ground state; when the wave function has a given number of nodes we can 
construct also the branches of solutions that -by analogy with the 
particle in a box- are called excited spherical states. However, as 
shown in \cite{GuzmanUrena2004,GuzmanUrena2006} such excited states are unstable 
and decay into ground states; in fact ground states are stable and late-time 
attractors for quite arbitrary initial wave function profiles 
\cite{GuzmanUrena2006,BernalGuzman2006}. Because the excited states are 
unstable and decay in a short time scale we collide in the present task only 
ground state configurations.

Once we account with these ground state data: i) we interpolated the wave 
function of such configuration centered at $(0,z_0)$, ii) we place another 
of these equilibrium configuration at the point $(0,-z_0)$, iii) we 
choose $z_0$ so that the two configurations are far 
enough one from the other (see below) and iv) resolved Poisson equation 
(\ref{eq:poisson}). Then we have initial data for two ground state equilibrium 
configurations in our axially symmetric domain.

Summarizing, we choose to solve the initial value problem in spherical 
coordinates to make sure that we start the evolution with very
accurate values, and we evolve in a 2D grid using cylindrical coordinates 
because we found them necessary and practical for the binary case 
we are interested in. 

\begin{figure}[htp]
\includegraphics[width=4cm]{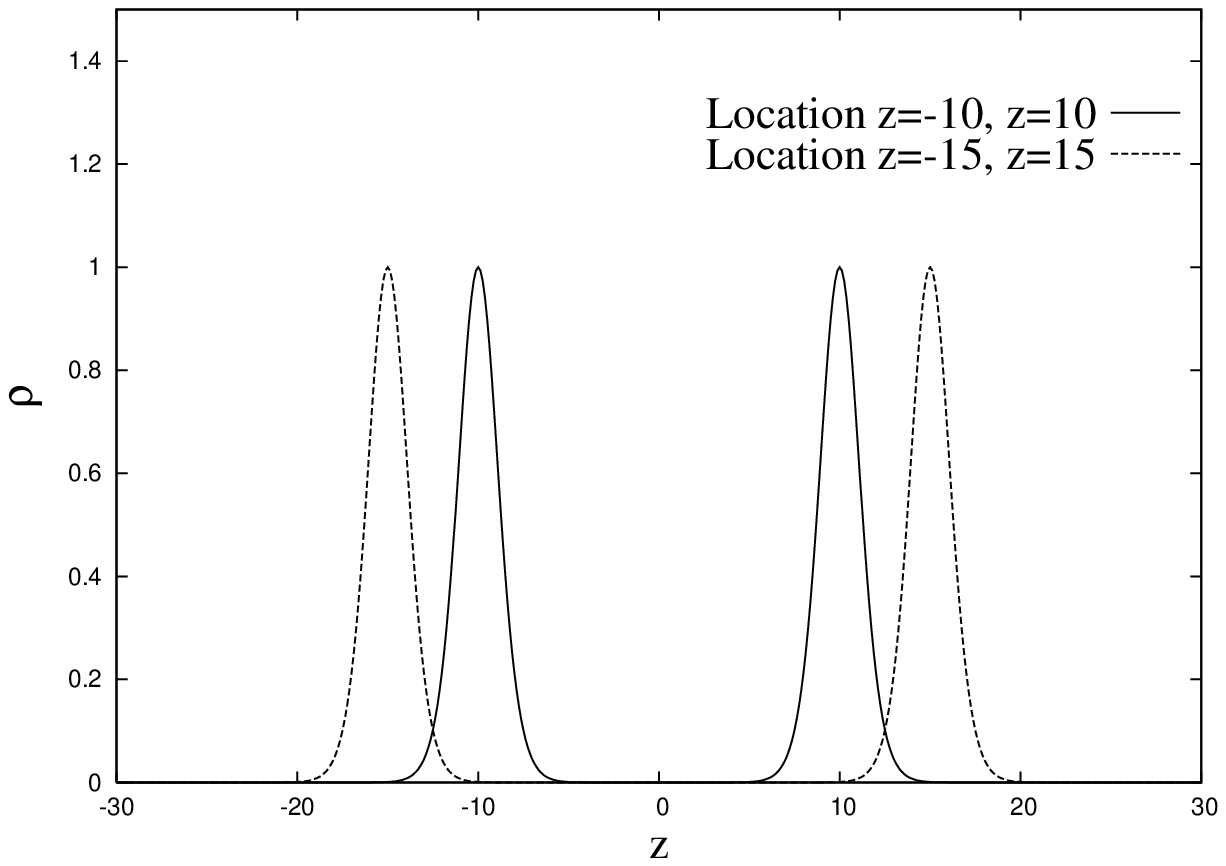}
\includegraphics[width=4cm]{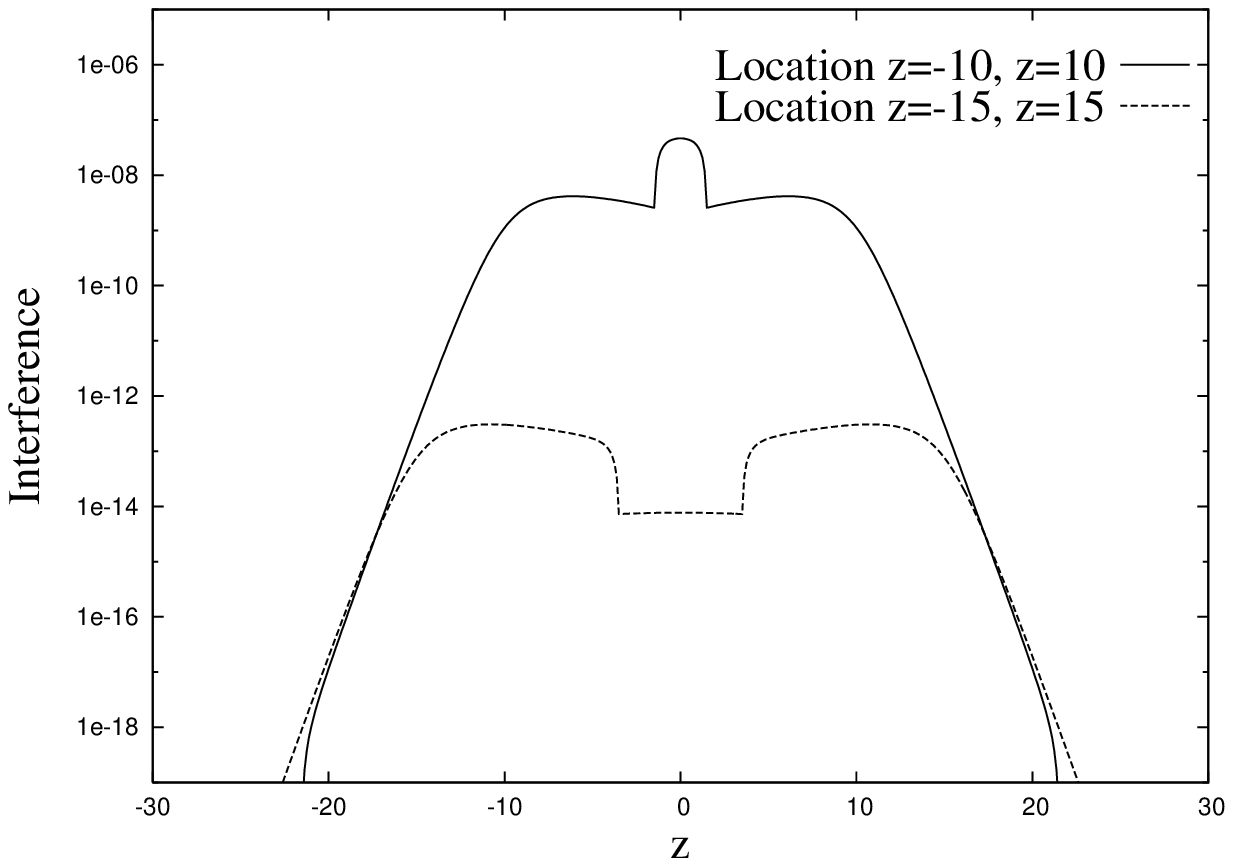}
\caption{\label{fig:ivp} (Left) An example of two ground state 
configurations superposed along the $z-$axis with two different 
separations. (Right) The interference $<\psi_1,\psi_2>$ is 
shown for two cases. Different separations are used and the
two configurations have a central field value $\psi(0)=1.0$. The initial 
head-on momentum is $p_z=3.0$.}
\end{figure}

Special warning is needed in Eq. (\ref{eq:schroedinger}), because it is 
non-linear for the $\Lambda \ne 0$ case, which indicates that the 
superposition of two wave functions is not allowed. Assume $\psi_1$ and 
$\psi_2$ represent the solutions of the initial spherically symmetric 
configurations that are to be superposed onto the 2D grid; the density 
of probability in (\ref{eq:poisson}) for the total wave function 
$\psi = \psi_1 + \psi_2$ is $|\psi_1 + \psi_2|^2$ and unless 
these states are orthogonal one cannot consider the naive superposition 
above is allowed. Thus we choose the distance between the configurations 
such that the interference (given by the scalar product of the 
two wave functions) $<\psi_1,\psi_2>$ is of the order of the 
precision of our calculations, say, in our case, the precision of the 
interpolation of the data into the 2D grid. Thus we can think of the 
system as one made of two adequately superposed equilibrium 
configurations we want to collide. An example of the interference term is shown 
in Fig. \ref{fig:ivp}.

The superposition of configurations by itself would say little about 
whether or not two configurations collide. We add an extra 
ingredient to the system, that is, an initial head-on momentum to the 
initial scalar field balls. We simply generate different initial 
kinematical states by assigning new values to the wave functions of the 
equilibrium configurations: $\psi_1 \rightarrow \psi_1 e^{ip_z z}$ 
and $\psi_2 \rightarrow \psi_2 e^{-ip_z z}$. The resulting physical 
situation involves a considerable change in the value of the 
expectation value of kinetic energy in the system.

Fortunately, in the present Newtonian low energy regime it is possible to 
estimate expectation values of physical quantities, a property difficult 
to pose in the fully general relativistic case. For instance, we account 
with the observables that allow one to monitor the evolution of the 
physical situation. Because we deal with a quantum mechanical system, we 
simply estimate the expectation values of the following interesting 
operators:

\begin{eqnarray}
K &=& -\frac{1}{2}\int \psi^{\ast} \nabla^2 \psi d^3x\\
W &=& \frac{1}{2}\int \psi^{\ast}U\psi d^3x\\
I &=& \int \Lambda |\psi|^4 d^3x
\end{eqnarray}

\noindent which are the expectation values of the kinetic, gravitational 
and self-interaction energies. These quantities are quite important at 
determining the state of the system at any time during the evolution of 
the system. That is, the value of the total energy $E = K+W+I$ 
indicates whether we account with a bounded system or not, and the very 
important virial theorem relation $2K+W+3I=0$ \cite{Wang2001}, which is 
nearly satisfied when the system gets virialized and relaxed through 
whatever channels available, for instance, the emission of scalar field 
bursts, the so called gravitational cooling.

% ----->     HEAD-ON COLLISIONS

\section{Head-on collisions and solitonic behavior}
\label{sec:headon}

% ->
\subsection{Equal mass case}

\begin{figure}[htp]
\includegraphics[width=8cm]{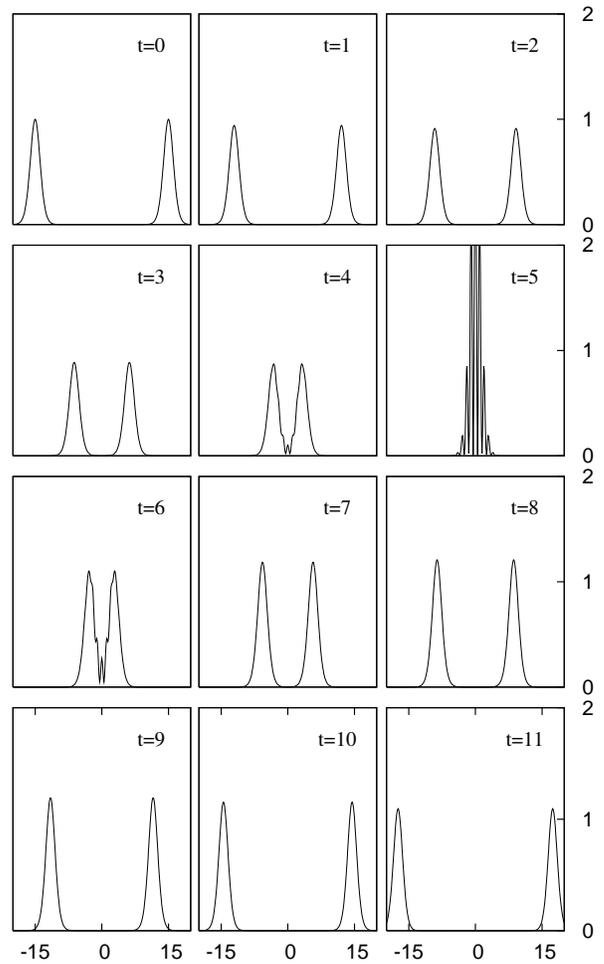}
\caption{\label{fig:snapshots} Snapshots of the density of probability for  
interaction between two initial 
configurations with $\Lambda=0$, $p_z=3.0$ and $z_0=15$. The configuration 
shows solitonic behavior due to the fact that the total energy is always 
positive of the order of $E \sim 32$ until the blobs get absorbed by the 
sponge. The numerical domain used is $x \in [0,30]$, $z \in [-30,30]$ with 
resolution $\Delta x = \Delta z = 0.125$.} 
\end{figure}

\begin{figure}[h]
\includegraphics[width=8cm]{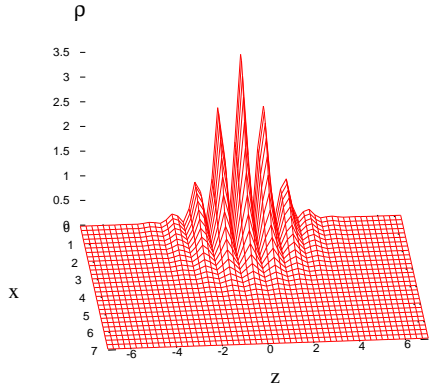}
\caption{\label{fig:pattern} The pattern of interference formed during 
the collision of two configuration with $\Lambda=0$, $z_0=15$, $p_z=3.0$
(the same case as in the previous figure). 
After this stage, the two blobs continue their way in the initial 
direction and behave as solitons (see Fig. \ref{fig:snapshots}). They 
are not strict solitons because they slightly deform during the process.} 
\end{figure}

% ----------------------
The first scenario one might think of is the collision of two equal mass 
ground state configurations. In Fig. \ref{fig:snapshots} we 
show snapshots of the density profile along the $z$-axis for an initial 
configuration with $p_z=3.0$ and $z_0=15$ in the free field case 
($\Lambda = 0$). What 
is found is that the two blobs move toward each other and eventually 
they lie upon each other, an 
interference pattern gets formed and after a 
while one blob moves toward the left and the other one toward the right. 
The first interpretation is that the initial data behave like solitons. 
Unfortunately we cannot be confident about the solitonic behavior 
because the shape of the blobs gets deformed after the ``collision'' 
and an increase of the amplitude and shrink in the width are manifest. 
After the distributions approach the boundary (located at $z=\pm 30$) 
the density of probability is absorbed by the sponge and its integral 
$M$ drops to zero. At this point we are unable 
to track the evolution further in time and we ignore whether the blobs 
might return and collide again and repeat such process as many times as 
desired until there is energy released (e.g. through the emission of 
scalar field) and the encounters get damped allowing eventually a true 
collision. In Fig. \ref{fig:pattern} we show a zoom of the interference 
pattern at the time when superposition of the configurations around 
$t \sim 5$ occurs.

\begin{figure}[htp]
\includegraphics[width=8cm]{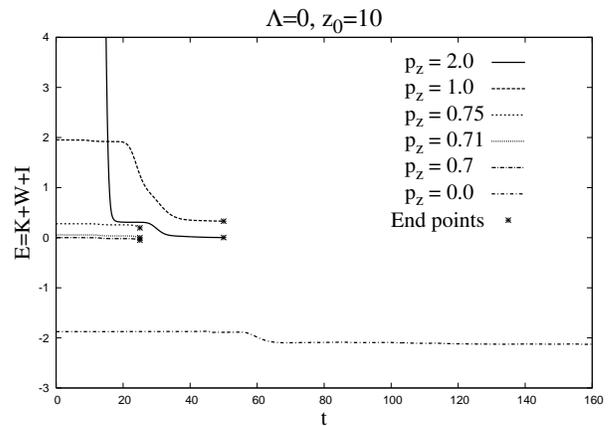}
\caption{\label{fig:energyL0.0} The total energy $E=K+W+I$ for different 
initial parameters, all of them with $\Lambda=0$ and $z_0=10$. The 
configurations with $p_z=2.0,1.0$ start evolving with high values of 
the total energy, and show solitonic behavior; the total energy tends to 
zero once the density of probability (the partciles) gets out of the numerical 
domain, which happens by the time indicated with the star for such run. 
The configuration with $p_z=0$ remains with clearly negative total 
energy, so that the system is bound and the system collides, in the 
sense that there is no solitonic behavior and instead the two blobs get 
glued and remain like that. We are unable to conclude anything about 
the borderline cases. The stars at the end of borderline cases 
indicate the point at which we stoped the runs. The reason is that the 
lenghtscale of such cases is pretty much that of our numerical domain, and one 
expects the truning points of the blobs to be at a distance of the order of the 
domain size. By the time indicated with a star, there is burst of particles, 
related more to the fact that the blobs are returning bach to the domain than to 
a burst of particles due to the relaxiation of a single blob or to the fact 
that the blobs are leaving completely the domain.}
\end{figure}

\begin{figure}[htp]
\includegraphics[width=8cm]{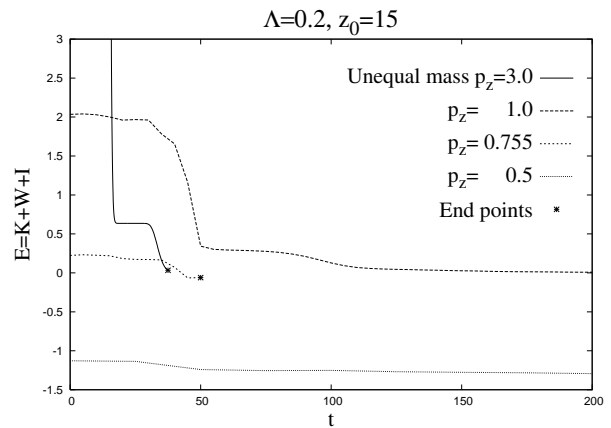}
\caption{\label{fig:energyL0.2} The total energy $E=K+W+I$ for different 
initial parameters, all of them with $\Lambda=0.2$ and $z_0=15$. 
Once again, the configuration with a clearly positive total energy 
($p_z=1.0$ and the unequal mass case with $p_z=3.0$) shows solitonic 
behavior and those with clearly negative total energy ($p_z \le 0.5$) show a 
merger. Once againg, we stopped the run of the borderline case with $p_z=0.755$, 
which physical properties change at about $t \sim 50$, and the numerical domain 
used does not suffice to determine its fate.}
\end{figure}

Of course, not all the initial configurations constructed present this 
behavior and we have found that a criterion to decide 
whether this behavior is allowed or not is the value of the total energy 
$E=K+W+I$. In Figs. \ref{fig:energyL0.0} and \ref{fig:energyL0.2} 
we show the total energy of different types of initial configurations. 
In Fig. \ref{fig:energyL0.0} we present different situations for the 
free field case $\Lambda=0$ and two particular cases: $p_z=1.0$ and 
$p_z=0$; in the first case the solitonic behavior is achieved and 
the total energy is always positive and approaches zero 
because the density of probability has left the numerical domain; in the 
second case the total energy is always negative and at the end of the 
day what is found is that there is a single blob in the middle, 
indicating that the system is oscillating around a bounded object. We 
show snapshots of this behavior later when dealing with the more 
interesting unequal mass case. About the other cases in this plot 
$p_z=0.75,0.71,0.7$ we cannot decide whether they show solitonic 
behavior or not in the time we used to run our simulations and we can 
only observe that the density profiles are severely distorted by the 
collision.

In Fig. \ref{fig:energyL0.2} we show the same criterion for 
configurations with the self-interaction term ($\Lambda=0.2$). Again, 
when the total energy is extremely positive or extremely negative we are able to 
decide whether the configurations collide or they trespass each other.

% ----------------------

% ->
\subsection{Unequal mass case}

The unequal mass case helps at deciding whether the configurations in 
the above examples truly trespass each other or bounce. In fact up to 
now it is not possible to say anything about this because of a few 
reasons: i) the expectation absolute value of the linear momentum along 
the head-on direction is equal for both half planes $z > 0$ and $z<0$, 
ii) the mass in each half plane is also the same in both 
semiplanes, iii) the expectation value of the linear momentum in the 
head-on direction is zero all the time.

\begin{figure}[htp]
\includegraphics[width=8cm]{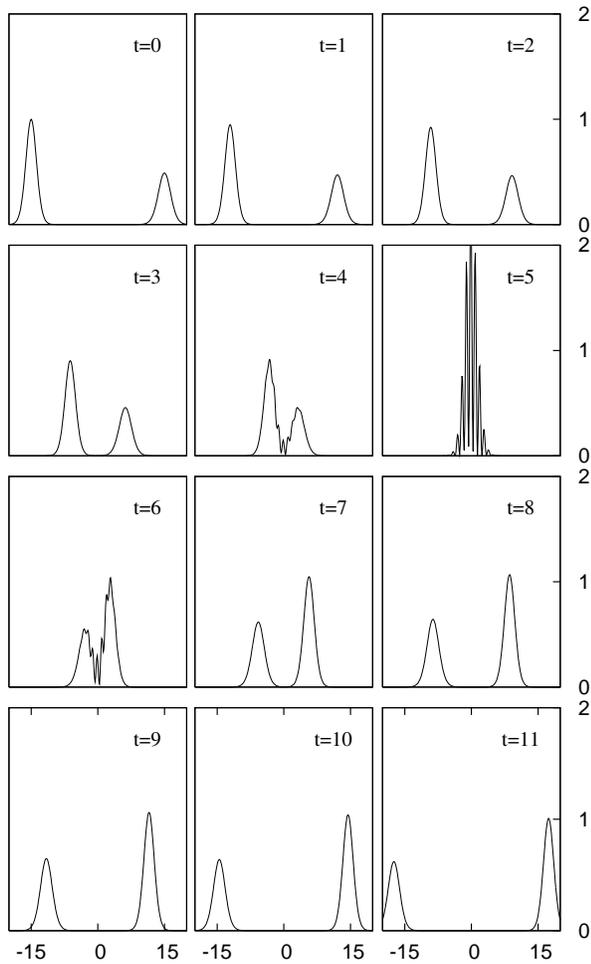}
\caption{\label{fig:snapshots-unequal} Snapshots of the density of probability 
for the interaction between two 
unequal mass initial ground state configurations with $\Lambda=0.2$, 
$p_z=3.0$ and $z_0=15$ (the first case in the previous figure). The 
configuration shows solitonic behavior due to the fact that the total 
energy is always positive of the order of $E \sim 35$ until the blobs 
get absorbed by the sponge. The total energy remains positive all the 
time, until the blobs are absorbed by the sponge. The numerical domain used is 
$x \in [0,30]$ and $z \in [-30,30]$ with resolution $\Delta x = \Delta z = 
0.125$.} 
\end{figure}

In the unequal mass case we have the advantage of being able to 
distinguish the half planes masses and linear momentum. In Fig. 
\ref{fig:snapshots-unequal} we show snapshots of the unequal mass case 
for $\Lambda=0.2$ and initial parameters $p_z=3.0$ and $z_0=15$ that 
show solitonic behavior with $E > 0$ all the way. It can be seen 
clearly that the initial blobs are actually trespassing each other 
although they suffer a profile deformation until the wave function 
reaches the sponge region.

\begin{figure}[htp]
\includegraphics[width=8cm]{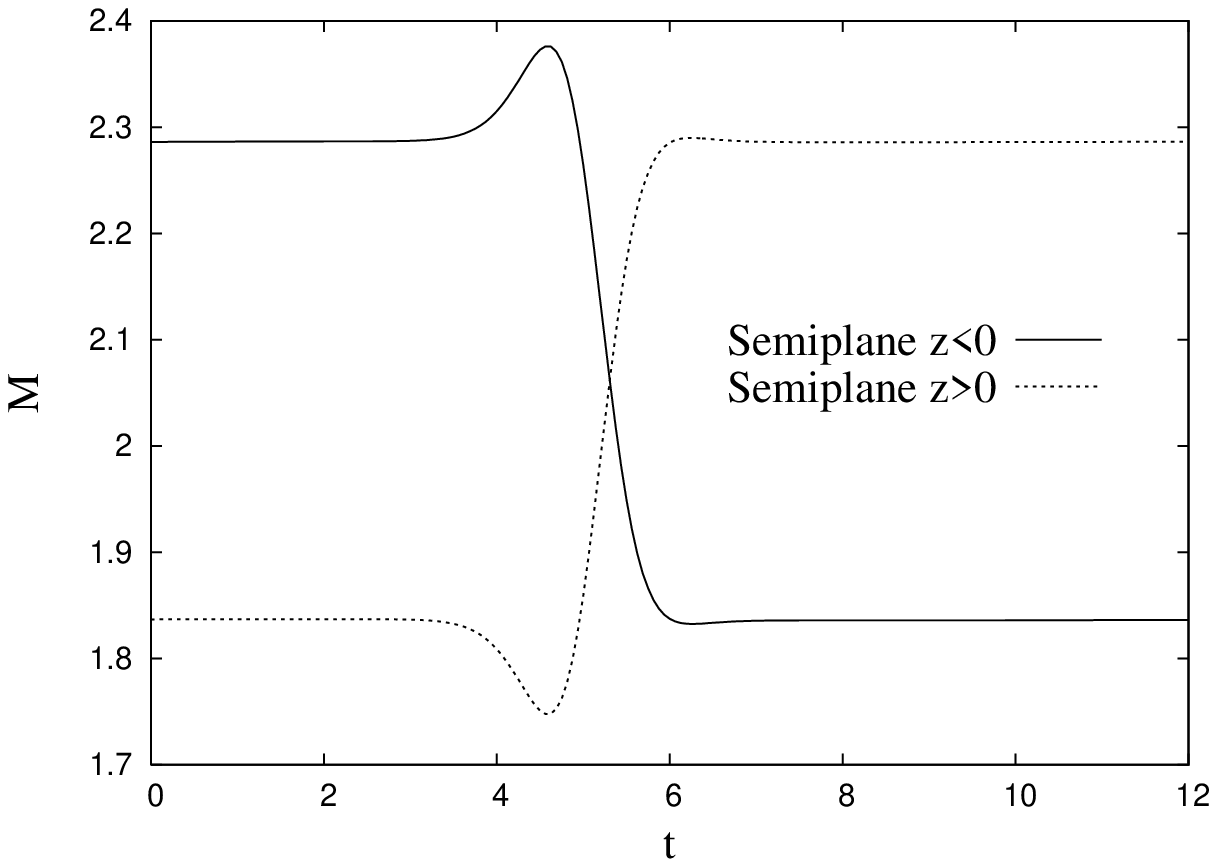}
\includegraphics[width=8cm]{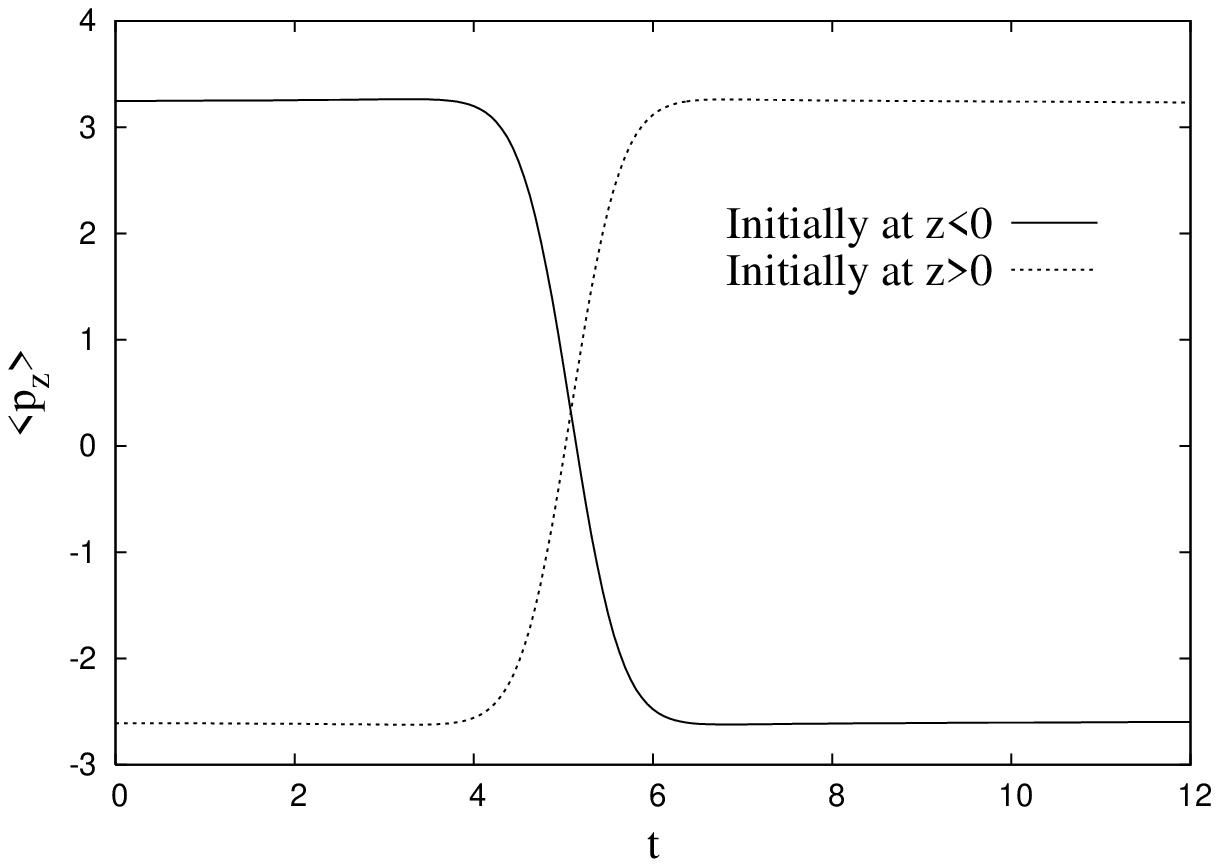}
\caption{\label{fig:unequal-scalars} Top: we show the transfer of 
particles from the $z<0$ half plane to the $z>0$ one and viceversa. 
The mass in each semiplane is shown. Bottom: we show the 
transfer of momentum between the two half planes, actually the 
expectation value calculated in each semiplane. The configuration 
evolved consists of two superposed ground state configurations with 
$\Lambda=0.2$, and respective central field values $\psi(0,0) = 1,0.7$ 
and masses $M=2.437,1.9383$. The initial linear momentum is $p_z = 3.0$. 
The density of probability reaches the edges of the numerical domain and 
vanishes. The total energy is always positive.} 
\end{figure}

In Fig. \ref{fig:unequal-scalars} we show the mass transfer from the 
$z>0$ to the $z<0$ half planes and viceversa; notice that the mass 
transfers from one side to the other in a very effective 
way. We also show the expectation value of the linear momentum along $z$ 
in both half planes; what is found is that the momentum is also 
transfered from one side to the other. The perfect solitonic behavior 
would consist of having these two properties plus the unachieved one 
related to the preservation of the density profile.

\subsection{An example of collision}

\begin{figure}[htp]
\includegraphics[width=8cm]{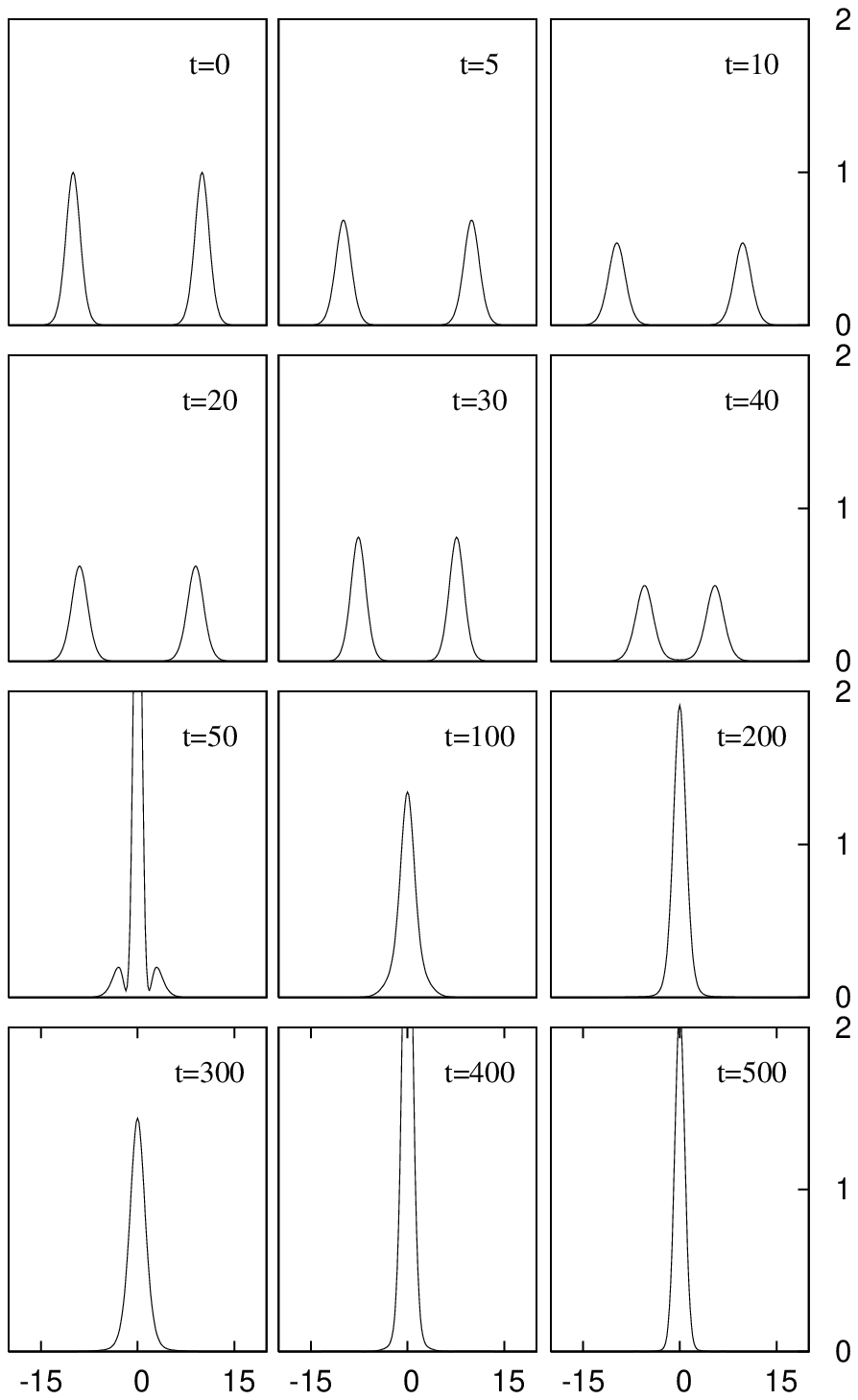}
\includegraphics[width=8cm]{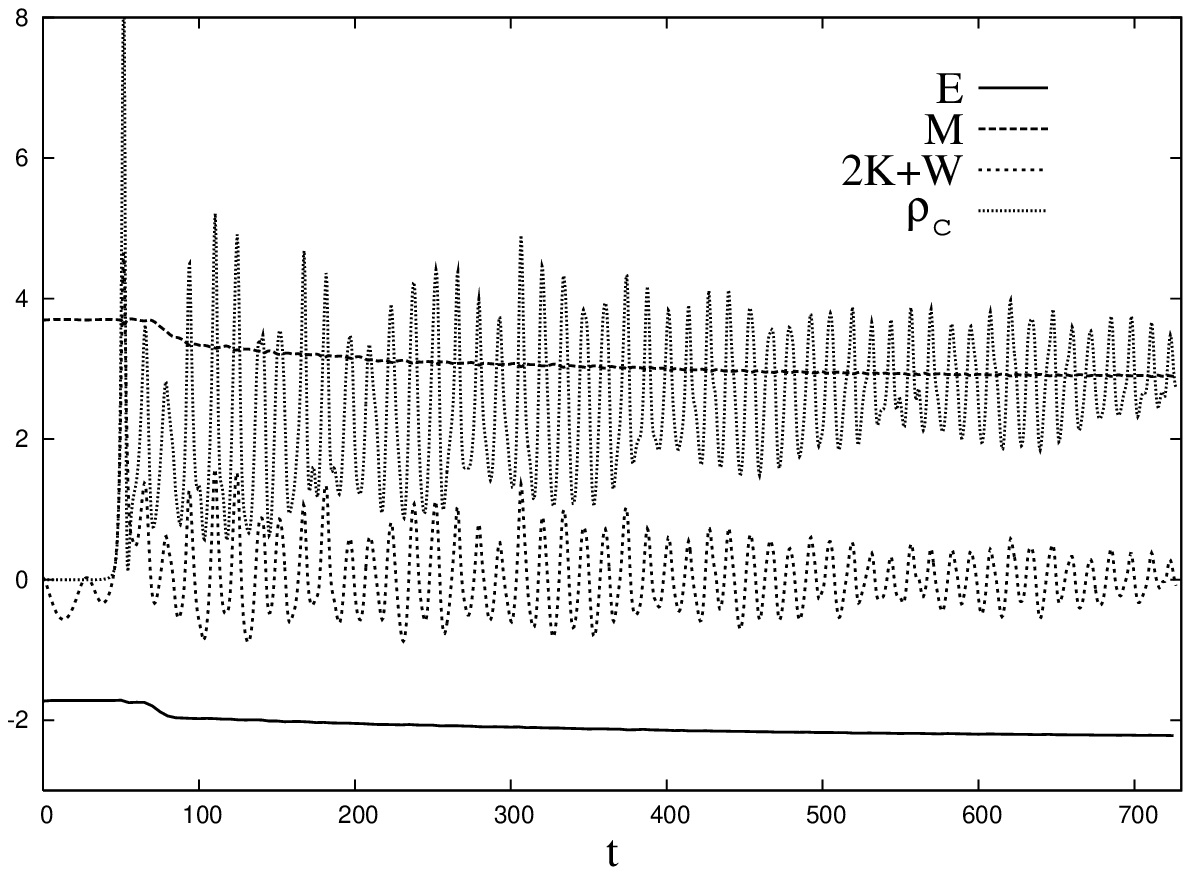}
\caption{\label{fig:snapshots-collision} Top: snapshots of the density of 
probability for an equal 
mass case with $\Lambda=0$ and initial parameters $p_z = 0$, 
$z_0=10$. The system apparently is trapped within the gravitational 
potential well. Bottom: the central density, total energy, mass, and virial 
relation are shown. The run-time is not enough to decide whether
the configuration will get virialized or not. However it behaves as a
perturbed system: the total energy is approaching a constant value
after a period of considerable activity, and the central density
starts oscillating around a given value.}
\end{figure}

As a final result we show what happens when an initial configuration 
presents a negative total energy. In Fig. \ref{fig:snapshots-collision} 
we show the density profile along $z$ for a collision case corresponding 
to $\Lambda=0$, $z_0=10$ and $p_z=0$, that is, only the pure gravitational force 
drives the dynamics of the binary. It can be observed that the 
blobs actually merge and remain sitting on a fixed point around the 
center of mass, the density tends to get stabilized, the virial relation starts 
oscillating aorund zero with smaller amplitude, the total energy starts 
stabilizing, so as the mass of the system.

In order to ullistrate what our results mean in physical units we use the run in 
Fig. \ref{fig:snapshots-collision} to estimate the time-scale for the collision 
of binary equal mass head-on case. We start from the fact that the mass of 
ground state configurations is $M \sim 10^{11}M_{\odot}$ and the mass of the 
boson is $m = 10^{-23}$eV; the separation is $20 \sim 3.52$kpc and the time the 
density peak is maximum is $t_{collision} \sim 52.5 \sim 8.3 \times 10^{6}yr$; 
the maximum relative speed before the collision is $v \sim 830 km/s$.

% ----->     CONCLUSIONS

\section{Conclusions}
\label{sec:conclusions}

We have presented numerical solutions of the Schr\"odinger-Poisson 
system of equations which includes the non-linear term related to the 
self-interaction in the mean field Gross-Pitaevskii equation for Bose 
Condensates. In such case, the potential well is given by self-gravity 
of the density of probability of the system. The particular case we have 
studied corresponds to the interaction between two ground state 
configurations (spherical both of them) 
\cite{GuzmanUrena2006}. 

We found that the initial blobs show solitonic behavior of the initial 
configuration, but that also the two configurations may collide. The 
system ends up colliding whenever the total energy of the system 
$E < 0$ and the solitonic behavior appears when $E > 0$. 
Unfortunately we can show this only for clearly non-zero values of the 
energy in each case and we cannot conclude anything about the borderline 
case, that is, when $E \sim 0$, because our simulations are unable to 
resolve the system for the quite long time needed and the spatial domain used.

Within the scalar field dark matter paradigm, the two initial blobs 
would represent two virialized structures made of dark matter. What we 
have shown is that not all couples of configurations are allowed to 
have a collision, and that the total energy would indicate whether or 
not a collision can occur. Our calculations also involve the presence 
of a self-interaction term in the scalar field, and therefore are 
within the Gross-Pitaevskii frame of Bose Condensates, which this time 
are gravitating.

% ----->     ACKNOWLEDGMENTS

\section*{Acknowledgments}

This research is partly supported by projects CIC-UMSNH-4.9 and 
PROMEP-UMICH-PTC-121. The runs were carried out in the Ek-bek cluster of 
the ``Laboratorio de Superc\'omputo Astrof\'{\i}sico (LASUMA)'' at 
CINVESTAV-IPN. A. B. acknowledges support from CONACyT.

% -------------------------------------
% 	      BIBLIOGRAPHY
% -------------------------------------

% ----->     END     <-----

\end{document}